\documentclass[%
 aip,
 amsmath,amssymb,
 reprint,%
]{revtex4-1}

\usepackage{graphicx}
\usepackage{dcolumn}
\usepackage{bm}

\usepackage[utf8]{inputenc}
\usepackage[T1]{fontenc}
\usepackage{mathptmx}
\usepackage{etoolbox}
\usepackage{xcolor}

\makeatletter
\def\@email#1#2{%
 \endgroup
 \patchcmd{\titleblock@produce}
  {\frontmatter@RRAPformat}
  {\frontmatter@RRAPformat{\produce@RRAP{*#1\href{mailto:#2}{#2}}}\frontmatter@RRAPformat}
  {}{}
}%
\makeatother
\begin{document}

\preprint{AIP/123-QED}

\title{Reversal of the Anomalous Hall Effect by Directional In-Plane Electric Field in Strained Bilayer $\text{MoO}$}
\author{Wenpeng Wang}
\author{Hang Shi}
\author{Xiangru Kong*}
\email{kongxiangru@mail.neu.edu.cn}
\affiliation{College of Sciences, Northeastern University, Shenyang 110819, China}

\date{\today}

\begin{abstract}
Layered magnetic materials with tunable valley and spin degrees of freedom hold promise for multifunctional spintronic devices. Here, we demonstrate that the orientation of an in-plane electric field enables selective switching of single-valley spin splitting and reversal of the anomalous Hall conductivity in strained bilayer $\mathrm{MoO}$. In particular, electric fields along the $a$ and $b$ directions selectively lift the spin degeneracy at the $X$ and $Y$ valleys, respectively, while preserving it at the other valley. This valley-selective switching is accompanied by a redistribution of Berry curvature with opposite signs, giving rise to a reversible anomalous Hall response. These effects are realized in the energetically and dynamically stable AC2 stacking configuration, where biaxial strain induces Dirac-like band crossings that enable subsequent electric-field control. Meanwhile, the AC1 stacking configuration, which is also energetically and dynamically stable, exhibits rich strain-driven phenomena, including a transition from metal to topological insulator with quantized spin Hall conductivity, as well as valley polarization and piezomagnetism. Our findings reveal a combined effect of strain and electric fields for controlling valley-selective spin splitting and anomalous Hall responses, highlighting bilayer $\mathrm{MoO}$ as a promising platform for multifunctional spintronic and valleytronic devices.
\end{abstract}

\maketitle

Two-dimensional (2D) magnetic materials offer opportunities to manipulate coupled spin, valley, and topological degrees of freedom~\cite{gibertini_magnetic_2019,burch_magnetism_2018,schaibley_valleytronics_2016,song_altermagnets_2025,PhysRevLett.133.166701,chen_unconventional_2025,9346-9jpf}. In particular, inequivalent valleys can exhibit distinct energies, Berry curvatures, and spin textures~\cite{PhysRevLett.89.226805,luo_valleytronics_2024,cao_valley-selective_2012,PhysRevLett.97.186404,gauswami_valley_2026,rycerz_valley_2007}, enabling valley-dependent transport phenomena such as the valley Hall effect and electrically tunable valley polarization~\cite{doi:10.1126/science.1250140,mv8w-7svx,isberg_generation_2013,wang_ferroelectricity_2026,PhysRevLett.99.236809,PhysRevLett.108.196802}. At the same time, magnetic order can further modify the symmetries protecting valley degeneracy~\cite{zhou_multiple_2024,zhou_symmetry_2024,gauswami_valley_2026}, making simultaneous spin and valley control attractive for spintronic and valleytronic applications~\cite{PhysRevLett.108.196802,PhysRevLett.115.037602,PhysRevB.105.045426}.


Beyond conventional valley polarization, selective spin splitting at one valley while preserving degeneracy at another offers a route to independent spin and valley control. Although valley polarization and spin–valley coupling have been widely studied~\cite{schaibley_valleytronics_2016,xu_spin_2014,PhysRevB.101.115423,tlpw-95yc,6wl8-r9zw}, realizing such selective splitting in compensated magnets remains challenging. Therefore, identifying the underlying symmetry conditions and external control mechanisms is therefore important for spin–valley filtering applications.

The monolayer $\text{MoO}$ has recently been proposed as a 2D altermagnet that combines compensated magnetic moments with momentum-dependent spin splitting~\cite{PhysRevB.107.214419}. Its valley-dependent electronic structure is sensitive to the underlying crystal symmetry, making it a suitable candidate for studying externally controlled valley and spin responses. However, the properties of an isolated monolayer are mainly determined by its intrinsic lattice and magnetic symmetry, which limits the available control parameters. Extending monolayer $\text{MoO}$ to a bilayer structure introduces additional degrees of freedom associated with stacking configurations and interlayer magnetic coupling, offering greater flexibility in tailoring magnetic symmetry and manipulating valley-dependent electronic properties through strain and electric fields~\cite{w468-mdrb,gauswami_valley_2026}.

Using first-principles calculations, we investigate the stacking-dependent magnetic, electronic, topological, and valleytronic properties of bilayer $\text{MoO}$. Among four stacking configurations, AC1 and AC2 are energetically and dynamically stable, exhibiting altermagnetic and compensated antiferromagnetic ordering, respectively. In AC1, biaxial tensile strain induces a transition from a metallic to a topological insulating state with quantized spin Hall conductivity, while uniaxial strain generates valley polarization and a piezomagnetic response. In AC2, biaxial tensile strain produces symmetry-related Dirac-like crossings, and an in-plane electric field enables valley-selective spin splitting, with the field orientation determining which valley exhibits spin splitting. These findings demonstrate the combined roles of stacking, strain, and electric fields in controlling magnetic, topological, and valley-dependent electronic properties, highlighting bilayer $\text{MoO}$ as a promising platform for multifunctional spintronic and valleytronic applications.

\begin{figure}[ht]
    \centering
    \includegraphics[width=0.49\textwidth]{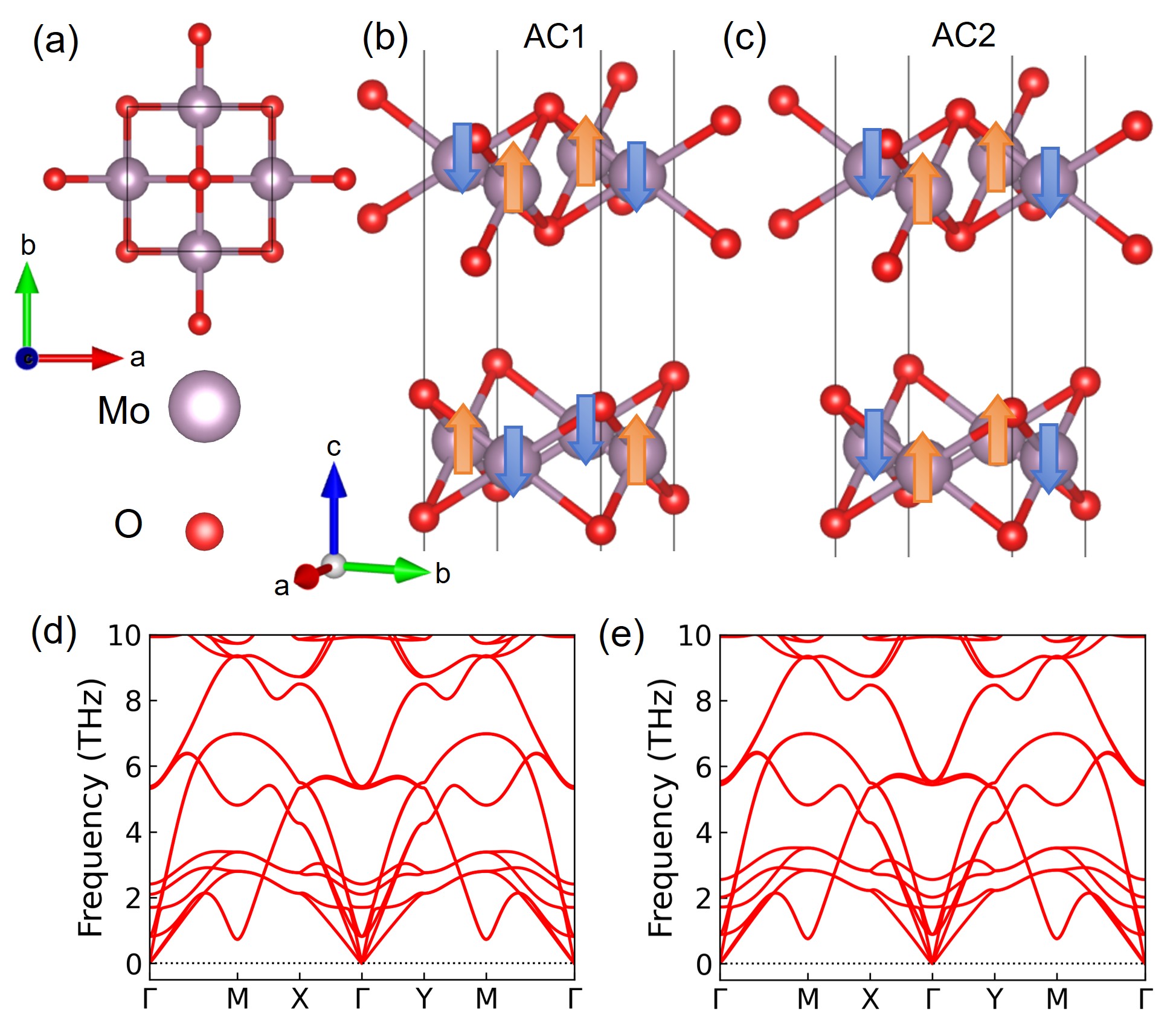}
    \caption{(a) Top view of AC-stacked bilayer $\text{Mo}\text{O}$. Magnetic configurations of (b) AC1 and (c) AC2, with red and blue arrows indicating opposite Mo magnetic moments. Phonon spectra of (d) AC1 and (e) AC2.}
    \label{fig:1}
\end{figure}

The monolayer $\text{Mo}\text{O}$ forms a typical sandwich structure: a single Mo atomic sheet lies between O sheets in a symmetric fashion. In Fig.~\ref{fig:1}(a), the oxygen atoms and molybdenum atoms are marked as red and pink, respectively. By stacking two layers of $\text{Mo}\text{O}$, four stacking configurations can be generated by considering four possible sliding directions, namely AA, AB, AC, and AD, which can be seen in Fig.~S1. Besides, the electronic band structures of the four stacking configurations are presented in Fig.~S2. We first investigate the structural stability and magnetic configurations of bilayer $\text{Mo}\text{O}$. For each stacking pattern (AA, AB, AC and AD), both parallel and antiparallel interlayer magnetic configurations are considered, and denoted as AA1/AA2, AB1/AB2, AC1/AC2 and AD1/AD2, respectively, with 1 and 2 indicating the parallel and antiparallel magnetic moment configurations. The total-energy comparison, shown in Table S1, shows that AC1 and AC2 are the two most energetically favorable configurations. As depicted in Fig.~\ref{fig:1}(d-e), the phonon spectrum shows no imaginary frequencies, confirming that the AC $\text{Mo}\text{O}$ bilayer is dynamically stable.

Both AC1 and AC2 preserve the same tetragonal parent crystal symmetry, characterized by the space group $P4/nmm$ and the crystallographic point group $D_{4h}$. Nevertheless, as shown in Fig.~\ref{fig:1}(b-c), their different collinear antiferromagnetic configurations lead to distinct magnetic space groups that $P_A^{\prime}/nm^{\prime}m$ for AC1 and $P4^{\prime}n^{\prime}m^{\prime}m$ for AC2. Thus, these two configurations exhibit distinctly different magnetic symmetries and electronic band structures. The magnetic symmetry of AC1 is consistent with an altermagnetic phase and allows a nonrelativistic spin splitting with a $d$-wave momentum dependence. By contrast, AC2 is a conventional collinear antiferromagnet, which contains combined $\mathcal{PT}$ symmetry that enforces twofold degeneracy at each $\mathbf{k}$ point in the Brillouin zone. Accordingly, no spin splitting or symmetry-allowed spin texture is expected for AC2, either without or with SOC. 

\begin{figure}[ht]
    \centering
    \includegraphics[width=0.49\textwidth]{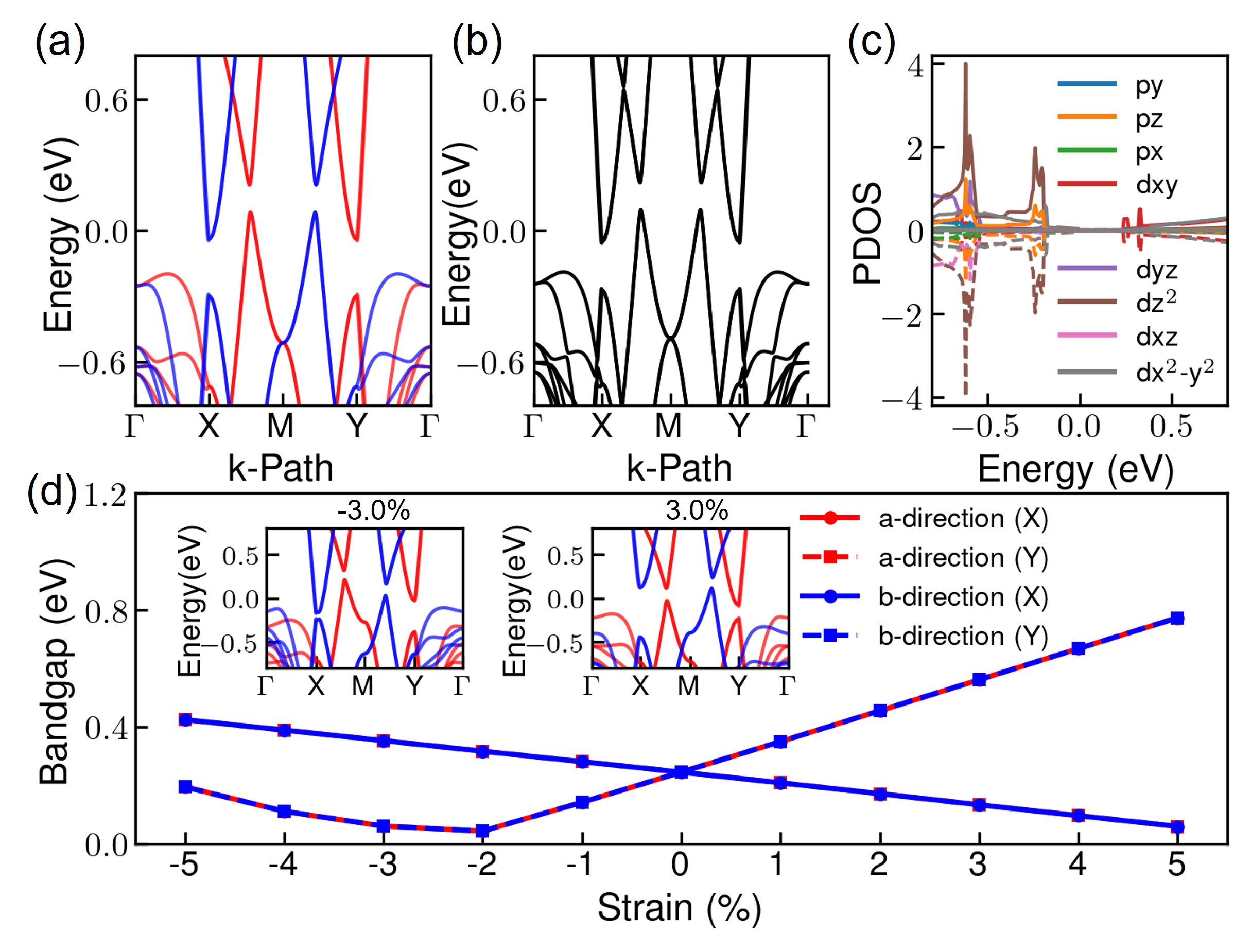}
    \caption{Electronic band structures of AC1 bilayer $\text{MoO}$ (a) without and (b) with SOC. (c) Spin-resolved PDOS, with positive and negative values denoting spin-up and spin-down contributions. (d) Strain-dependent band gaps at the \(X\) and \(Y\) valleys. Insets show the band structures under \(3\%\) compressive and tensile strain along the \(a\) direction.
    }
    \label{fig:2}
\end{figure}

In the absence of external strain, AC1 is metallic because one band cross the Fermi level. This metallic character is robust against the inclusion of SOC, which can be seen in Fig.~\ref{fig:2}(a) and (b). The spin-resolved band structure further reveals pronounced momentum-dependent spin splitting near the Fermi level. In Fig.~\ref{fig:2}(c), orbital-projected band analysis shows that the splitting is predominantly associated with Mo-\(d_{z2/xy}\) and O-\(p_z\) orbitals.

\begin{figure}[ht]
    \centering
    \includegraphics[width=0.5\textwidth]{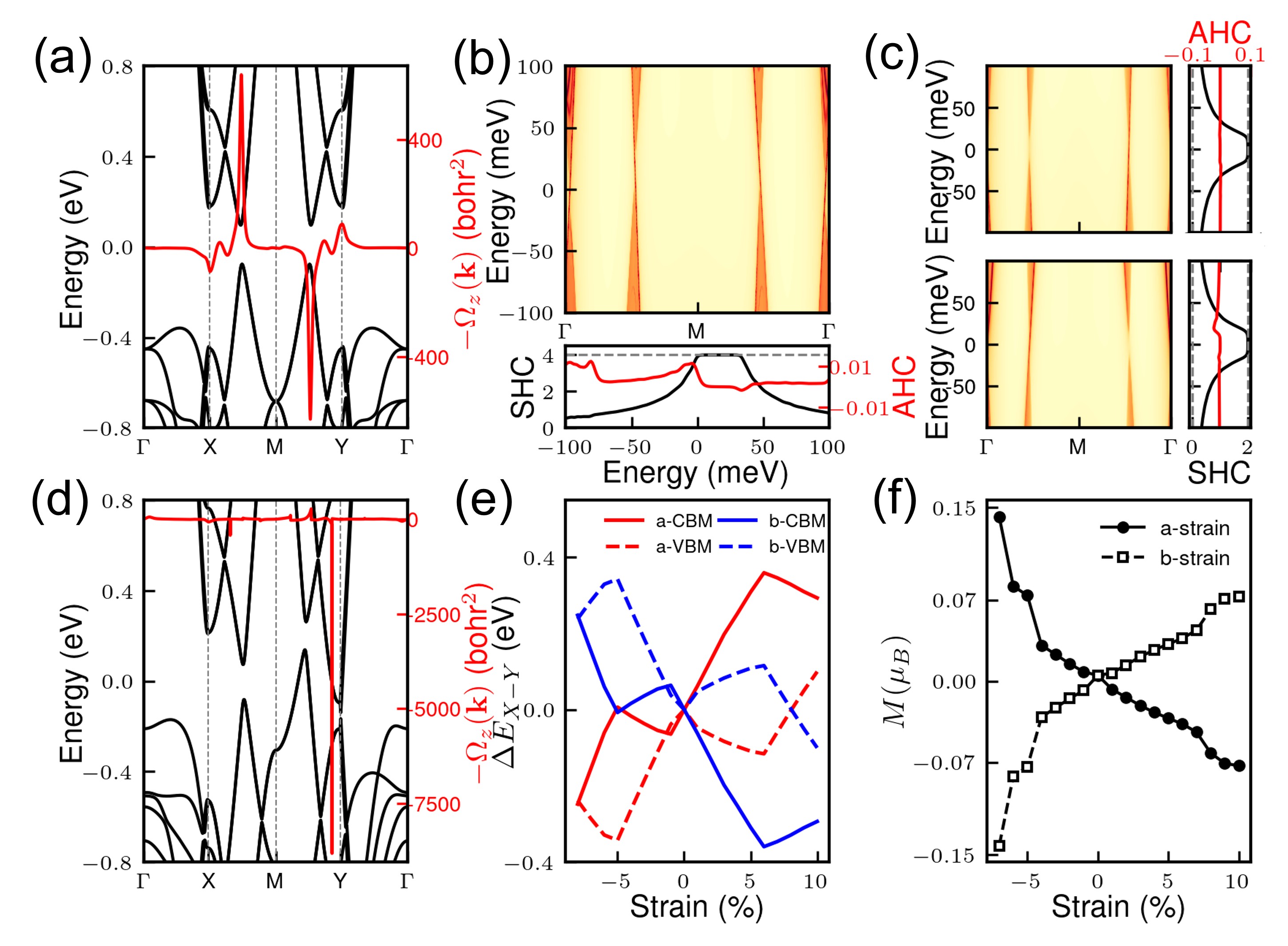}
    \caption{Topological and strain-induced properties of AC1 bilayer $\text{Mo}\text{O}$. (a) Band structure and Berry curvature under \(5\%\) biaxial tensile strain. (b) Edge states, SHC, and AHC of the bilayer. (c) Corresponding layer-resolved results for the upper and lower layers. (d) Band structure under \(5\%\) uniaxial tensile strain along the \(a\) direction. (e) Strain-dependent valley polarization in the conduction (solid) and valence (dashed) bands. (f) Net magnetization versus uniaxial strain.
    }
    \label{fig:3}
\end{figure}

Tensile and compressive strains affect physical quantities such as lattice bond lengths differently, thus they can produce distinct responses in AC1. Compressive strain preserves the metallic character, whereas tensile strain induces distinct changes in the electronic band structure. Under uniaxial tensile strain, one valley remains metallic because a band continues to cross the Fermi level, while a local band gap opens at the other valley. This state can be referred to as a half-valley metal (Fig.~\ref{fig:3}(d)). The valley-selective electronic response depends sensitively on the direction of uniaxial strain. To quantify the valley polarization, we define the conduction- and valence-band valley splittings as
\(\Delta E_{c/v}=E_{c/v}\left(X\right)-E_{c/v}\left(Y\right)\), separately for the conduction-band minimum (CBM) and valence-band maximum (VBM). As shown in Fig.~\ref{fig:2}(b), \(\Delta E_{c/v}\) is nearly zero in the unstrained AC1 bilayer, indicating that the \(X\) and \(Y\) valleys remain symmetry equivalent. Uniaxial strain breaks the fourfold rotational symmetry that relates the two valleys and consequently lifts their valley degeneracy.
Uniaxial strain breaks the fourfold rotational symmetry relating the \(X\) and \(Y\) valleys, lifting their degeneracy. As shown in Fig.~\ref{fig:3}(e), the valley polarization of conduction band changes sign from compressive to tensile strain along the \(a\) direction, while strain along \(b\) produces the opposite trend. The valley polarization of valence band also exhibits sign reversals, demonstrating that both strain direction and magnitude can control the relative valley energies. The evolution of the electronic band structures under uniaxial strain applied along the \(a\) and \(b\) directions is presented in Fig.~S3. Therefore, uniaxial strain provides an efficient means to control the valley polarization and selectively favor distinct valleys in AC1 bilayer $\text{Mo}\text{O}$.

In addition, uniaxial strain induces a finite net magnetization in AC1 bilayer $\text{Mo}\text{O}$. As illustrated in Fig.~\ref{fig:3}(f), the unstrained system exhibits a vanishing net magnetization due to the complete compensation of the magnetic moments on the two sublattices, as enforced by the $S_{4z}\mathcal{T}$ symmetry. Upon applying uniaxial strain along the \(a\) or \(b\) direction, this symmetry is broken, allowing a finite net magnetization to emerge. Remarkably, the induced magnetization exhibits opposite responses to strain along the two crystallographic directions. Specifically, as the strain changes from compressive to tensile, the net magnetic moment reverses from positive to negative for strain along the \(a\) direction, while the opposite trend is observed for strain along the \(b\) direction. These results reveal an anisotropic piezomagnetic response arising from the coupling between lattice deformation and magnetic order. In particular, the ability to control both the magnitude and sign of the magnetization without an external magnetic field may be advantageous for developing strain-controlled spintronic functionalities and magnetic sensing devices~\cite{ikhlas_piezomagnetic_2022}.

Compared with uniaxial strain that breaks the original crystal symmetry, biaxial tensile strain modifies interatomic distances and electronic hopping strengths without breaking the underlying crystal symmetry~\cite{https://doi.org/10.1002/adfm.202302330,wu_strain-reshaped_2026}. These modifications lead to a substantial reconstruction of the electronic band structure near the Fermi level and may open a finite band gap. As a result, the evolution of the band structures under biaxial strain (Fig. S4) demonstrates that the AC1 bilayer $\text{Mo}\text{O}$ undergoes a transition from a metallic state to a globally insulating phase. The evolution of the band gap under tensile and compressive biaxial strains exhibits a pronounced dependence on both the strain magnitude and the specific valley, as illustrated by the band structures in the inset of Fig.~\ref{fig:2}(d). At a tensile strain of approximately $5\%$, the system develops a global band gap of $0.191$~eV, which can be seen in Fig.~\ref{fig:3}(a). We first compute the Berry curvature in momentum space. As shown in Fig.~\ref{fig:3}(a), pronounced Berry curvature hot spots emerge in the regions related to $M$ - $X$ and $M$ - $Y$ of the Brillouin zone. These Berry curvature hot spots have nearly equal magnitudes but opposite signs. The opposite Berry curvatures at the $X$ and $Y$ valleys are protected by the surviving antiunitary $C_{4z}\mathcal{T}$ symmetry, which transforms one valley into the other and requires
$\Omega_z(X)=-\Omega_z(Y)$. 
To further examine the transport implications of the Berry curvature distribution, we calculate the anomalous Hall conductivity (AHC), which can be evaluated as~\cite{PhysRevLett.49.405}
\[
\sigma_{xy}^{\mathrm{A}}
=
-\frac{e^{2}}{\hbar}
\sum_{n}
\int_{\mathrm{BZ}}
\frac{d^{2}k}{(2\pi)^{2}}
f_{n\mathbf{k}}\,
\Omega_{n,z}(\mathbf{k})
\approx 0.
\]
As shown in Fig.~\ref{fig:3}(c), the entire bilayer system exhibits a vanishing AHC. Layer-resolved calculations of AHC further reveal negligible anomalous Hall responses in both the top and bottom $\text{Mo}\text{O}$ layers, indicating that the overall vanishing conductivity does not arise from the cancellation of finite, oppositely signed layer contributions.

In contrast to the anomalous Hall response, the spin Hall conductivity exhibits a quantized value. The intrinsic spin Hall conductivity is calculated as~\cite{PhysRevLett.95.156601}
\[
\sigma_{xy}^{S}
=
-\frac{e}{\hbar}
\sum_{n}
\int_{\mathrm{BZ}}
\frac{d^{2}k}{(2\pi)^{2}}
f_{n\mathbf{k}}\,
\Omega_{n,xy}^{s_z}(\mathbf{k}),
\]
where $\Omega_{n,xy}^{s_z}(\mathbf{k})$ denotes the spin Berry curvature associated with the $z$ component of the spin current. At the selected Fermi level, the total spin Hall conductivity reaches
\[
\sigma_{xy}^{S}
\approx
4\,\frac{e}{\hbar},
\]
which is approximately decomposed into equal contributions from the top and bottom $\text{Mo}\text{O}$ layers,
\[
\sigma_{xy,\mathrm{top}}^{S}
\approx
\sigma_{xy,\mathrm{bottom}}^{S}
\approx
2\,\frac{e}{\hbar}.
\] Furthermore, the edge spectrum exhibits two co-propagating chiral edge branches on each edge, accompanied by the spin Hall conductivity of \(\frac{4e}{\hbar}\), as shown in Fig.~\ref{fig:3}(b). These results indicate that the two layers contribute constructively to the spin Hall response, whereas their anomalous Hall contributions are zero (Fig.~\ref{fig:3}(c)). 

\begin{figure}[ht]
    \centering
    \includegraphics[width=0.5\textwidth]{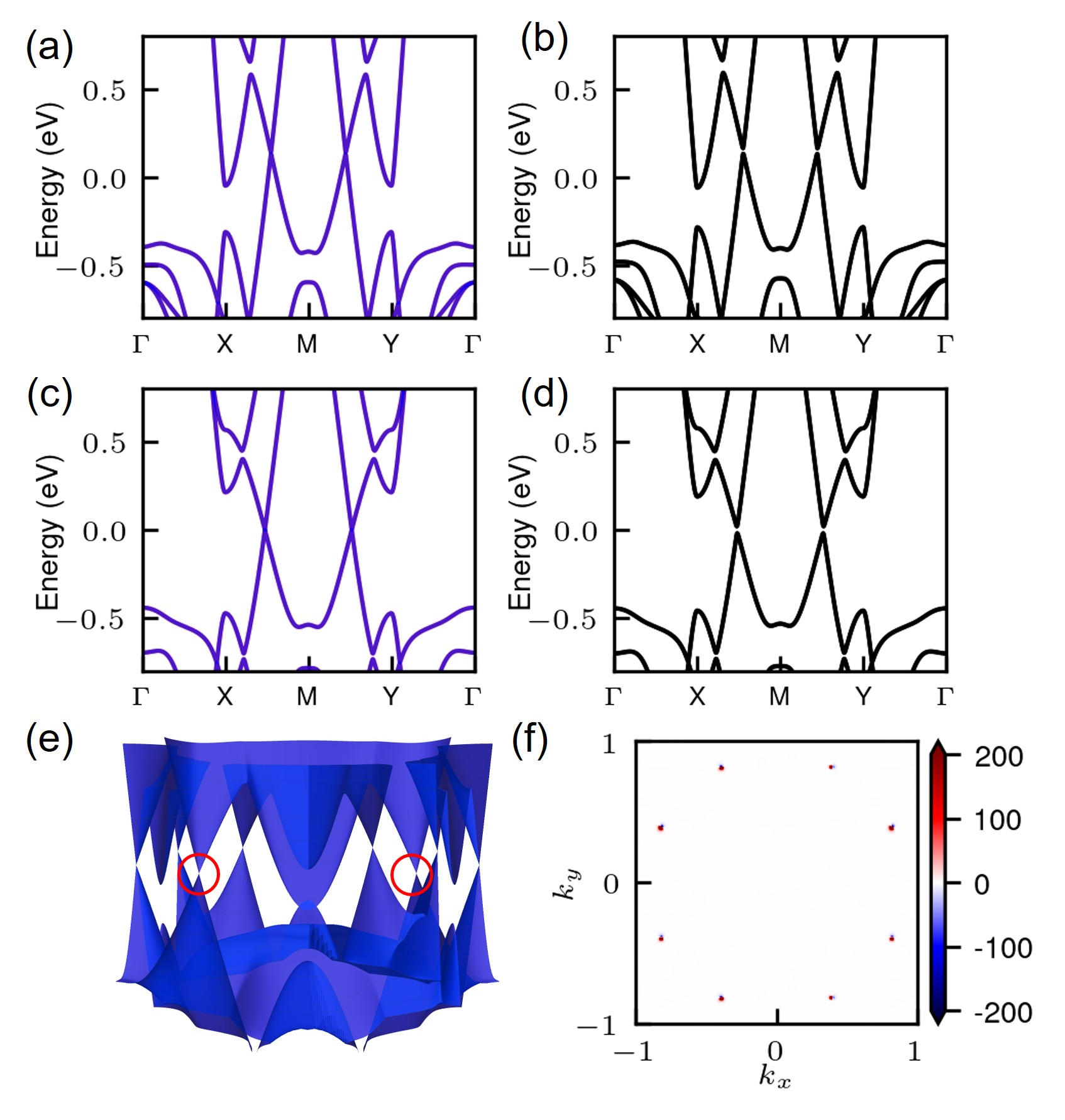}
    \caption{Strain-induced Dirac states in AC2 bilayer $\text{Mo}\text{O}$. Band structures (a,b) without strain and (c,d) under \(5\%\) biaxial tensile strain, calculated without (a,c) and with (b,d) SOC. (e) Three-dimensional energy dispersion and (f) Berry curvature distribution near the Dirac points under \(5\%\) biaxial tensile strain.
    }
    \label{fig:4}
\end{figure}

The AC2 configuration provides a distinct platform for exploring novel spin-dependent phenomena beyond those observed in AC1. In this configuration, the two layers possess opposite magnetic moments, forming a compensated antiferromagnetic state. Without external strain, the spin-up and spin-down bands remain degenerate in the Brillouin zone, giving rise to a metallic ground state, as illustrated in Fig.~\ref{fig:4}(a).


Biaxial tensile strain induces symmetry-protected band crossings along the $M$-$X$ and $M$-$Y$ paths, forming two symmetry-related Dirac valleys in AC2 bilayer $\text{Mo}\text{O}$ (Fig.~\ref{fig:4}(c-e) and S5). Including SOC opens a small gap of $\Delta_{\mathrm{SOC}}=0.042$ eV at these crossings (Fig.~\ref{fig:4}(d)). Besides, the Berry curvature distribution near the Dirac points in the AC2 $\text{Mo}\text{O}$ bilayer under \(5\%\) biaxial tensile strain is presented in Fig.~\ref{fig:4}(f). However, biaxial tensile strain alone cannot induce single-valley spin splitting. It reconstructs the electronic band structure of AC2 bilayer $\text{Mo}\text{O}$ near the $M$-$X$ and $M$-$Y$ directions, thus establishing the valley degrees of freedom required for subsequent single-valley spin splitting under electric field.

\begin{figure*}[ht]
    \centering
    \includegraphics[width=1.0\textwidth]{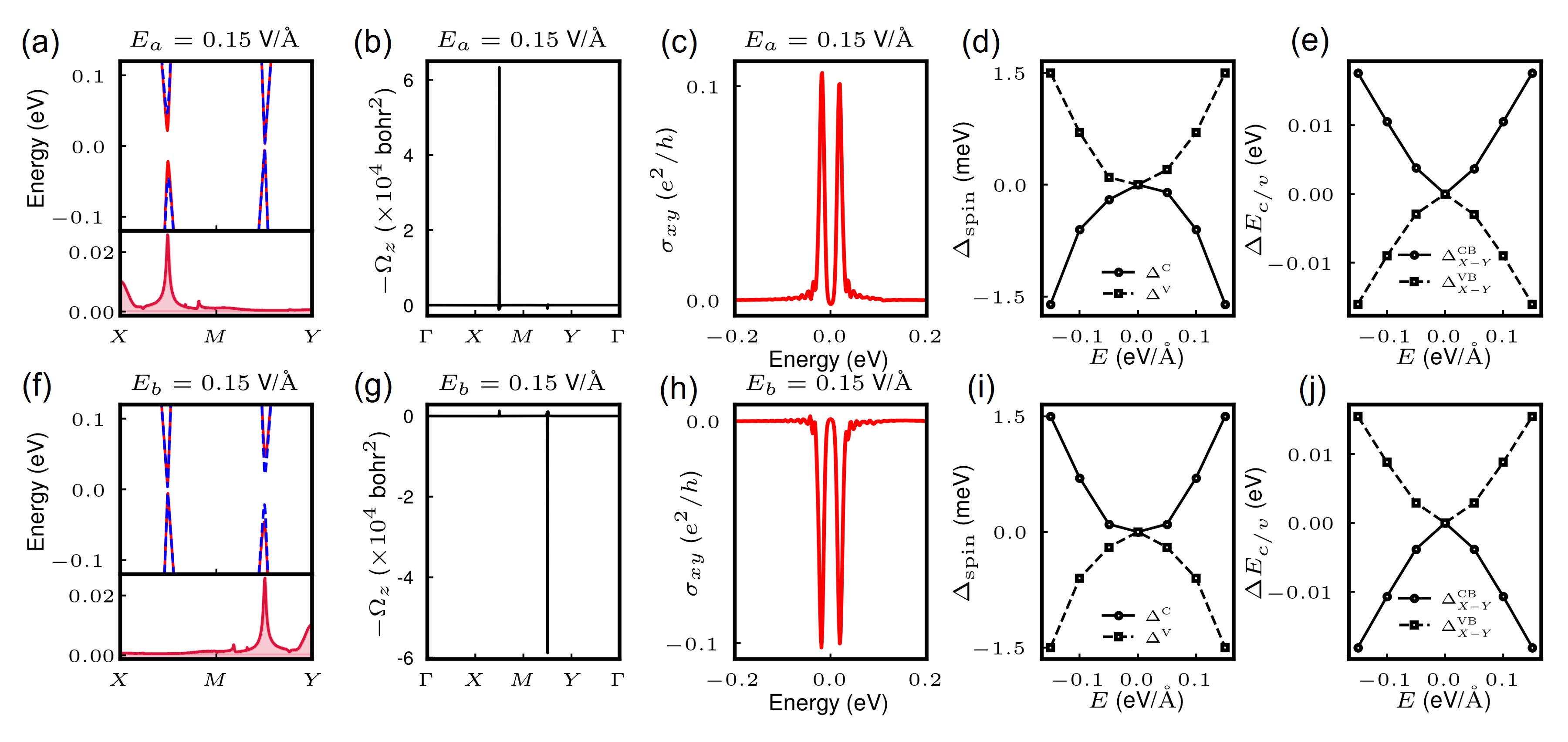}
    \caption{Direction-dependent valley-selective spin splitting and anomalous Hall responses in AC2 bilayer $\text{Mo}\text{O}$. (a,f) Spin-resolved band structures and spin-splitting magnitudes under electric fields along the \(a\) and \(b\) directions, respectively. (b,g) Corresponding Berry curvature distributions and (c,h) energy-dependent AHC. Electric field dependence of (d,i) conduction- and valence-band spin splitting and (e,j) valley polarization.}
    \label{fig:5}
\end{figure*}

Having established the role of strain in manipulating the valley properties, we further explore the combined effects of biaxial strain and an in-plane electric field on the valley degrees of freedom. The electric field response of AC2 bilayer $\text{Mo}\text{O}$ is strongly anisotropic. An electric field applied along the \(a\) direction selectively lifts the band degeneracy along the $M$-$X$ line, while the bands along $M$-$Y$ remain nearly degenerate. Here, the magnitude of the spin-splitting strength is defined as \[\left|\Delta_{\mathrm{spin}}\right|=\left|E_{up}-E_{down}\right|,\] as explicitly shown in the lower panels of the Fig.~\ref{fig:5}(a). As shown in the upper panels of the Fig.~\ref{fig:5}(a), a predominantly spin-up band splitting emerges along the $M$-$X$ path, whereas the bands along the orthogonal $M$-$Y$ path remain spin degenerate. Furthermore, \(\left|\Delta_{\mathrm{spin}}\right|\) reaches its maximum along the $M$-$X$ line. In contrast, when an electric field was applied along the \(b\) direction, a predominantly spin-down band splitting was induced along the $M$-$Y$ and \(\left|\Delta_{\mathrm{spin}}\right|\) reaches its maximum along the $M$-$Y$ line, while the $M$-$X$ bands remain nearly degenerate (Fig.~\ref{fig:5} (f)). The splitting induced by electric field evolves approximately with the electric field strength and exhibits a strong directional selectivity. As shown in Fig.~\ref{fig:5}(d) and (i), the spin splitting in both the conduction and valence bands increases with the magnitude of the electric field applied along the 
\(a\) and \(b\) direction, respectively. However, its magnitude is essentially independent of the direction of the in-plane electric field along a given crystallographic axis. This behavior suggests that reversing the direction of electric field reverses the sign of \(\Delta_{\mathrm{spin}}\), or equivalently exchanges the two spin channels, without changing \(\left|\Delta_{\mathrm{spin}}\right|\). 


When no external electric field is applied, AC2 is a compensated collinear antiferromagnet possessing $\mathcal{PT}$ symmetry. The antiunitary operation $\mathcal{PT}$ maps each state at a given $\mathbf{k}$ point onto an orthogonal degenerate partner at the same $\mathbf{k}$ point. Consequently, no global spin splitting occurs in the zero-field state. The magnetic space group contains the antiunitary symmetries $C_{2x}\mathcal{T}$ and $C_{2y}\mathcal{T}$, which leave the $M$-$Y$ and $M$-$X$ high-symmetry lines invariant, respectively, and impose additional constraints on the band representations along these lines. When an in-plane electric field with $\mathbf{E}_a\parallel a$ is applied, $C_{2x}\mathcal{T}$ is preserved, whereas $C_{2y}\mathcal{T}$ is broken. Consequently, the bands along the $M$-$Y$ line remain subject to the constraint imposed by the surviving symmetry, and the corresponding band splitting is suppressed. In contrast, the $M$-$X$ line loses the associated symmetry protection, thereby allowing spin splitting to emerge. For an electric field with $\mathbf{E}_b\parallel b$, the situation is interchanged between the $a$ and $b$ directions that $C_{2y}\mathcal{T}$ is preserved, whereas $C_{2x}\mathcal{T}$ is broken. As a result, a pronounced spin splitting develops along the $M$-$Y$ line, while the bands along the $M$-$X$ line remain approximately degenerate.



The in-plane electric field breaks the symmetry relating the $X$ and $Y$ valleys, lifting their energy degeneracy and inducing direction-dependent valley polarization. Specifically, electric fields applied along the $a$ and $b$ directions preferentially induce valley polarization at the $X$ and $Y$ valleys, respectively. To quantify this effect, we define the valley energy polarization for CBM or VBM as
\begin{equation}
\Delta_{X-Y}^{\mathrm{CB/VB}}
=
E_{X}^{\mathrm{CB/VB}}
-
E_{Y}^{\mathrm{CB/VB}},
\end{equation}
where $E_{X}^{\mathrm{CB/VB}}$ and $E_{Y}^{\mathrm{CB/VB}}$ are the corresponding CBM/VBM energies at the $X$ and $Y$ valleys, respectively. The evolution of $\Delta_{X-Y}^{\mathrm{CB/VB}}$ as a function of the strength of electric field is shown in Fig.~\ref{fig:5}(e) and (j), demonstrating the tunability of valley polarization through the direction and magnitude of the applied electric field.
This direction-dependent valley polarization is accompanied by a redistribution of the Berry curvature, which in turn modifies the anomalous Hall response. To elucidate this behavior, we calculate the Berry curvature and AHC of AC2 bilayer $\mathrm{MoO}$ under electric fields along the $a$ and $b$ directions. As shown in Fig.~\ref{fig:5}(b) and (g), the Berry curvature exhibits pronounced contributions of opposite signs along the $M$–$X$ and $M$–$Y$ paths, respectively, while remaining relatively weak along the other path in each case. Consequently, the corresponding AHC exhibit opposite signs near the Fermi level, as illustrated in Fig.~\ref{fig:5}(c) and (h). These results demonstrate that the orientation of the in-plane electric field enables selective control of valley polarization and the associated anomalous Hall response.

In summary, we have demonstrated the combined control of magnetic, topological, and valley-dependent properties in bilayer $\mathrm{MoO}$ through stacking, strain, and in-plane electric fields. The stable AC1 configuration exhibits altermagnetism, with biaxial tensile strain inducing a topological insulating state characterized by quantized spin Hall conductivity and vanishing AHC. Uniaxial strain further enables valley polarization and a piezomagnetic response. In the stable AC2 configuration, biaxial tensile strain generates symmetry-related Dirac valleys, while an in-plane electric field selectively lifts spin degeneracy at either the \(X\) or \(Y\) valley. Switching the field orientation redistributes the Berry curvature and reverses the sign of the anomalous Hall conductivity. These results highlight bilayer $\mathrm{MoO}$ as a platform for electrically tunable spin, valley, and anomalous Hall responses.

This work was supported by LiaoNing Revitalization Talents Program (Grant No. XLYCE2504021) and the
Fundamental Research Funds for the Central Universities (No. N25LPY025). X. K. acknowledges the start up funding from Northeastern University, China. 

There are no conflicts to declare.  

\bibliographystyle{unsrt} 
\bibliography{references} 



\end{document}